\documentclass[runningheads]{llncs}
\usepackage{bbding}
\usepackage{graphicx}
\usepackage[T1]{fontenc}
\usepackage{hyperref}
\hypersetup{
    colorlinks=true,       
    linkcolor=blue,        
    citecolor=blue,        
    urlcolor=blue          
}
\usepackage[linesnumbered, ruled]{algorithm2e}
\usepackage[noend]{algpseudocode}
\usepackage{pgfplots}
\pgfplotsset{compat=1.18}
\usepackage{pgfplotstable}
\usepackage{subfigure}
\usepackage{float}
\usepackage{amsfonts}
\usepackage{listings} 
\usepackage{amssymb}
\usetikzlibrary{patterns}
\usetikzlibrary{patterns.meta}

\newcommand{\eat}[1]{}
\newcommand{\ie}{\emph{i.e.,}\xspace}
\newcommand{\eg}{\emph{e.g.,}\xspace}
\newcommand{\aka}{\emph{a.k.a.,}\xspace}

\begin{document}
\title{SLSM : An  Efficient Strategy  for Lazy Schema Migration on Shared-Nothing Databases}
\titlerunning{SLSM}
%
\author{Zhilin Zeng\inst{1} \and
Hui Li\inst{1} \and
Xiyue Gao\inst{1(}\Envelope\inst{)} \and
Hui Zhang\inst{2} \and
Huiquan Zhang\inst{1} \and
Jiangtao Cui\inst{1}}
\authorrunning{Z. Zeng et al.}
%
\institute{
Xidian University, Xi'an, China \\
\email{zengzhilin.xidian@gmail.com, 20009200055@stu.xidian.edu.cn, \{hli,xygao,cuijt\}@xidian.edu.cn} \and
Shanghai Yunxi Technology Co., Ltd., Shanghai, China\\
\email{zhanghui@inspur.com}}
\maketitle              
\begin{abstract}
By introducing intermediate states for metadata changes and ensuring that at most two versions of metadata exist in the cluster at the same time, shared-nothing databases are capable of making online, asynchronous schema changes. However, this method leads to delays in the deployment of new schemas since it requires waiting for massive data backfill. To shorten the service vacuum period before the new schema is available, this paper proposes a strategy named SLSM for zero-downtime schema migration on shared-nothing databases. Based on the lazy migration of stand-alone databases, SLSM keeps the old and new schemas with the same data distribution, reducing the node communication overhead of executing migration transactions for shared-nothing databases. Further, SLSM combines migration transactions with user transactions by extending the distributed execution plan to allow the data involved in migration transactions to directly serve user transactions, greatly reducing the waiting time of user transactions. Experiments demonstrate that our strategy can greatly reduce the latency of user transactions and improve the efficiency of data migration compared to existing schemes.

\end{abstract}
\vspace{-3ex}\section{INTRODUCITON}
Due to the high scalability and availability, shared-nothing NewSQL systems like CockroachDB ~\cite{taft2020cockroachdb} and TiDB ~\cite{huang2020tidb} have been widely adopted in practice. As practical applications are constantly being updated and iterated, the database schema may need to evolve correspondingly. In fact, some studies have shown that schema changes occur on average once a week~\cite{qiu2013empirical}, while some cloud applications change the schema more frequently per week ~\cite{neamtiu2011cloud}. Schema changes, \aka schema migrations, are typically done with Data Definition Language (DDL) statements, and the process of schema migration can incur massive data movement that may block concurrent queries and transactions. It was common for early systems to require service downtime or maintenance windows to perform schema migration~\cite{dumitras2009no}. To avoid this, standard RDBMSs today use online schema change techniques or external migration tools~\cite{facebook2010,OAK2010,LHM2012,gh-ost2016,pt-osc2016,de2017zero,zhu2017towards}. The main strategy for these solutions is similar and usually involves snapshot replication and incremental data capture. Firstly, the system receives the DDL statements for the schema changes, which are processed to register a new schema and its corresponding table~\cite{facebook2010,OAK2010,LHM2012,gh-ost2016,pt-osc2016}, \aka a shadow table, and the shadow table is not yet available for service. Some approaches use materialized views instead of shadow tables ~\cite{de2017zero,zhu2017towards}. After that, the system copies the source table in the background (snapshot replication), \ie migrates the data from the old schema to the shadow table. Since the source table is not yet offline and is still available for service, when the system processes requests to change the data in the source table, these modified data need to be synchronized with the shadow table (incremental capture) to ensure consistency, which is achieved by means of triggers~\cite{facebook2010,OAK2010,LHM2012,pt-osc2016,de2017zero,zhu2017towards} bound on the source table or log propagation~\cite{gh-ost2016}. Once all data replication is complete, the system locks the source table and renames the shadow table to complete the atomic switch between the old and new schemas. The techniques described above only allow the new schema to serve after all the data in the source table have been synchronized, and schema changes can take more time if the old schema has a larger amount of data in the table.

Online schema change techniques in shared-nothing NewSQL databases follow the idea of mainstream RDBMS schemes and tend to introduce metadata transition states to ensure availability and consistency, as proposed in Google F1 ~\cite{rae2013online}. F1 introduces two additional intermediate metadata states, $delete\_only$ and $write\_only$, in addition to the two non-intermediate states $absent$ and $public$. After all F1-Servers have reached the $write\_only$ state, the database allows double writes to both old and new schema and simultaneously turns on backfill of new schema data. The data in the $delete\_only$ and previous phases and the data written in the $write\_only$ phase are generally referred to as stock and incremental data, respectively. This is the same as snapshot replication and incremental capture in the mindset of mainstream RDBMS solutions. Therefore, unfortunately, the flaws of mainstream RDBMS solutions are inherited along with it. However, as continuous deployment(CD) and integration became the norm ~\cite{laukkanen2017problems}, developers are expected to simply deploy updates to the front-end code and the back-end database. Waiting for the changes to the back-end database schema to be completed before deploying updates to the front-end code undoubtedly delays the service of the new schema. This prevents developers from realizing the benefits of CD as the effects need to be predicted and estimated prior to performing each schema change, reserving enough time to perform migration and locate rewrites for outdated queries in the source code. The cycle time to bring new schemas online is significantly lengthened.

Although some schemes adopting the lazy migration idea can avoid the delayed deployment of new schemas~\cite{bhattacherjee2021bullfrog,saur2016evolving,sheng2019non}, this idea has not yet been practiced on shared-nothing NewSQL databases. Inspired by the schemes in ~\cite{saur2016evolving,sheng2019non}, in this paper, we propose a zero-downtime Strategy for Lazy Schema Migration (SLSM) on shared-nothing databases. SLSM inherits the advantages of existing delayed migration schemes. More importantly, it is tailored to the architecture and data distribution characteristics of shared-nothing databases, greatly reducing the overhead of the system in executing migration transactions and user transactions. We conducted extensive experiments on one of the popular shared-nothing databases using standard TPC-C benchmarks that include schema migration transactions. According to our experimental results, our approach achieves a performance improvement of more than $40\%$ over existing solutions.

In summary, this paper makes the following contributions.
\begin{itemize}
    \item[1)] We propose and implement the SLSM, a lazy schema migration method on shared-nothing databases that supports single-step online schema migration.
    \item[2)] We reduce the overhead of SLSM in executing migration transactions and user transactions and optimize the background migration algorithm.
    \item[3)] We apply and integrate our approach to a popular shared-nothing database system and perform extensive experiments to demonstrate the advantages of our approach.
\end{itemize}
The rest of the paper is organized as follows. Section~\ref{sec:relwork} summarizes and reviews related work. Preliminaries are described in Section~\ref{sec:prel}. Section~\ref{sec:slsm} details the basic methodology and optimization principles of SLSM. The experimental results and evaluations are given in Section~\ref{sec:exp}. Finally, Section~\ref{sec:concl} summarizes this work.
\vspace{-3ex}\section{Related Works}\label{sec:relwork}
The idea of using lazy migration for nonblocking schema migration has been discussed in the standalone RDBMS ~\cite{bhattacherjee2021bullfrog,hu2022online,sheng2019non} and Nosql~\cite{saur2016evolving} database systems.

Sheng~\cite{sheng2019non} proposed a multi-stage lazy migration approach for adding columns, where the system logically adds a new schema when a schema change is performed without physically executing any data migration. Data from the same table may be physically stored in multiple tables under different schemas. The system handles queries by interpreting the data according to the most recent schema, moving the data from the old schema to the new one only when necessary. However, the method is strictly limited to one kind of schema change, \ie adding columns. Moreover, the characteristic of maintaining multiple physical tables increases the complexity and overhead of the system. Finally, the approach is discussed only in a standalone database system named Terrier and cannot be adapted to shared-nothing database systems.

BullFrog~\cite{bhattacherjee2021bullfrog} is an RDBMS that adopts a similar idea for schema change through lazy migration, with the difference that the system only needs to maintain two schema versions and the new schema begins working as soon as it is registered. When the system receives a request on a new schema, it migrates the relevant data involved in the old schema first and then processes the request on the new schema. This scheme supports a wider variety of schema change operations but is currently only available on PostgreSQL, a standalone RDBMS. \cite{saur2016evolving} starts with the single-step migration requirement and uses a lazy migration approach in the context of migrating an application on top of a NoSQL database (Redis). NoSQL databases are widely used by continuous deployment applications since they typically do not enforce schema constraints. For shared-nothing NewSQL databases, it is necessary to have both the features of an RDBMS and the scalability and high availability of a NoSQL system, and our work on SLSM demonstrates that lazy migration can also be well applied to distributed shared-nothing databases.

Tesseract~\cite{hu2022online} is a new method of online and transactional schema evolution. By modeling schema evolution as data modification and leveraging the concurrency control protocol, Tesseract can provide online, transactional schema evolution with no downtime and high application performance during the ongoing evolution progress. Experiments with Tesseract have demonstrated that it can work with existing lazy migration approaches to support the immediate deployment of schema changes. Our SLSM is compatible with Tesseract to provide a high-performance MVCC (Multi-Version Concurrency Control) scheme for non-blocking schema migrations on shared-nothing databases.

In contrast to previous work, SLSM provides a common solution for shared-nothing databases. It utilizes generic data partitioning rules and a simple modification of the \textsf{INSERT} operator for migration optimization without relying on a complex or ad-hoc design.
\vspace{-3ex}\section{Preliminaries}\label{sec:prel}
In this section, we introduce some preliminary knowledge related to the shared-nothing databases.
\vspace{-3ex}\subsection{Architecture for shared-nothing databases }\label{sec:architec}
In the classic shared-nothing~\cite{stonebraker1986case} architecture, a database cluster consists of an arbitrary number of nodes, located in the same data center or not. Clients are assumed to be able to connect to any node in the cluster. Individual nodes typically have a tiered architecture designed to optimize distributed data storage and transaction processing. Depending on the specific database implementation, tiering is not strictly defined or limited, but usually includes an SQL layer and a distributed storage layer.

The SQL layer follows the concept of traditional relational databases and is considered as an interface for users to interact with the database. The SQL layer is responsible for parsing the user's high-level SQL statements, generating distributed query plans, and converting them into low-level read and write requests to the underlying key-value store. 

The distributed storage layer slices the data using a range partitioning policy and is responsible for routing addressing of the slices to provide uniform KV storage. The data are ordered and automatically partitioned into small slices (called ranges or regions), with each slice containing a segment of contiguous key-value pairs. Tuples of tables are mapped into one or several key-value pairs, typically the key consists of its table ID and the primary key column, while the value is the actual row data. For example, a tuple containing four columns might be encoded as (where col0 is the primary key column):

$Key: \{tableID/col0\}$;     
$Value: \{col1/col2/col3\}$

Each slice has multiple replicas, and the replicas are replicated and maintained via a consistent consensus protocol (\eg raft~\cite{ongaro2014search} or paxos~\cite{lamport2001paxos}). Data of a certain table may be distributed on different nodes of the cluster according to range, and the system automatically performs load balancing and failover of the replicas to ensure high availability of the cluster.

\vspace{-3ex}\subsection{Raft Consensus Protocol}
For ease of understanding, we also describe the Raft protocol that is popular in many shared-nothing databases. In a cluster, all the replicas of a slice form a Raft group, where one replica is the long-lived \textit{leader} that coordinates all read and write operations sent to the Raft group, and the other replicas are the \textit{follower}. The raft protocol is usually used in conjunction with the lease mechanism. Only the lease-holding replica can provide consistent KV reads/writes for the slice. In addition, the replica holding the lease is usually the leader of the Raft group, and its node is the \textit{leaseholder}, typically a node may act as the leaseholder for multiple slices. Gateway nodes are responsible for parsing SQL requests, acting as transactions coordinators, and routing KV operations to the correct leaseholder. 
\vspace{-3ex}\section{SLSM Basic Methods and Optimization Principles}\label{sec:slsm}
In this section, we propose the basic SLSM scheme and further explore the optimization strategies. At the very beginning, we present a pair of definitions that are fundamental for the following discussion.
\begin{definition}
\textbf{User Transaction.} Obsolete queries on the old schema are rewritten as the new schema comes online. User transactions are those that contain requests on the new schema.
\end{definition}

\begin{definition}
\textbf{Migration Transaction.} Migration transactions are those transactions generated by and executed before user transactions. Migration transactions are responsible for migrating the old schema data involved in the user transactions to the new schema.
\end{definition}

\vspace{-3ex}\subsection{Basic Approach}
\begin{figure}[t]
\centering
\includegraphics[scale=0.6]{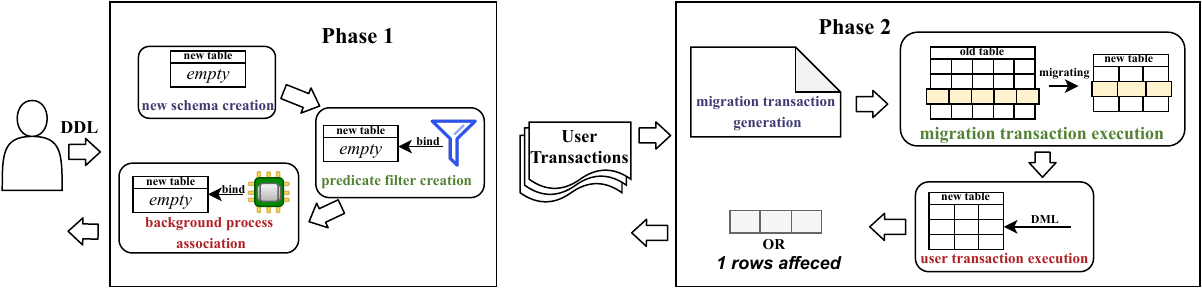}
\vspace{-2ex}\caption{Basic SLSM Overview}\label{overview}\vspace{-3ex}
\end{figure}

We first describe the basic approach of SLSM, where we modify the SQL engine layer of the shared-nothing database in two phases so that schema changes can be accomplished immediately without waiting for massive data to be migrated in place. We show the overall framework of SLSM in Fig.~\ref{overview}, The first phase of SLSM mainly handles schema change requests and performs the required initialization. The second phase mainly handles user requests on the new schema, where SLSM first migrates the data involved in the user query from the old schema and then processes the user request on the new schema. Logically, a given tuple starts in the old schema and eventually migrates to the new schema, but never exists in both schemas simultaneously. We shall elaborate on each step below.

\vspace{-3ex}\subsubsection{Initialization}
A schema migration request appears in the form of one or more DDL statements, \eg \textsf{CREATE TABLE new\_table ... AS SELECT...FROM old\_table}. Once the request is submitted, SLSM starts a transaction that creates an empty table corresponding to the new schema in the migration request and a predicate filter. For most shared-nothing databases that support view disambiguation techniques, the predicate filter can be replaced by a temporary view containing the contents of the migration request \eg \textsf{CREATE VIEW new\_table\_view ... AS SELECT...FROM old\_table}, otherwise some query rewriting tools may be required. SLSM then associates a background migration process with the new table to complete the initialization.

\vspace{-3ex}\subsubsection{Migration and User Transactions}
The initialization cost is trivial because no physical data migration is involved. Once initialization is complete, the new schema is ready for service. When the system receives a transaction request for the new schema from the user, \eg \textsf{SELECT, DELETE, UPDATE}, instead of executing the \textit{user transaction} immediately, SLSM first generates and executes a \textit{migration transaction}. The migration transaction is responsible for migrating any relevant data from the old schema by using a predicate filter to convert the filter predicates on the new schema in the user's transaction (usually located in the \textsf{WHERE} clause) to predicates on the old schema, forming a conditional \textsf{SELECT} statement on the old schema. As a result, it in fact acts as a subquery of the \textsf{INSERT} statement used by the migration transaction, such as \textsf{INSERT INTO new\_table (...) (SELECT ... from old\_table WHERE ...)}. After executing the migration transaction, the new schema has the set of tuples needed to fully process the user transaction, and the SLSM can then process the original user transaction request. It is worth mentioning that if the user transaction is a \textsf{INSERT} request, there is no need to perform a migration transaction, which can be handled directly on the new schema.
\vspace{-3ex}\subsubsection{Background Migrations}
SLSM starts a background migration process that slowly injects simulated user transactions that cumulatively overwrite the entire old table with tuples that have not yet been migrated. The end of the background migration thread indicates the completion of the migration, and the old schema tables can be deleted.

\vspace{-3ex}\subsection{Further Optimization for SLSM}\label{sec:furOpt}
\begin{figure}[t]
\centering
\includegraphics[scale=0.5,height=3cm]{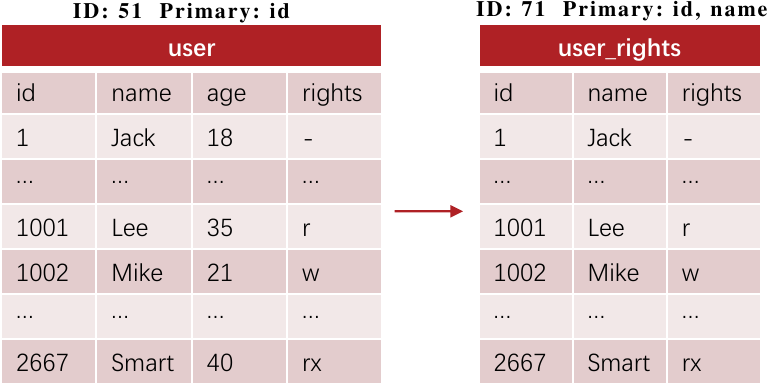}
\vspace{-2ex}\caption{Table Split Migration}\label{fig4.0}\vspace{-3ex}
\end{figure}

\begin{figure}[t]
\centering\vspace{-3ex}
\includegraphics[scale=0.22]{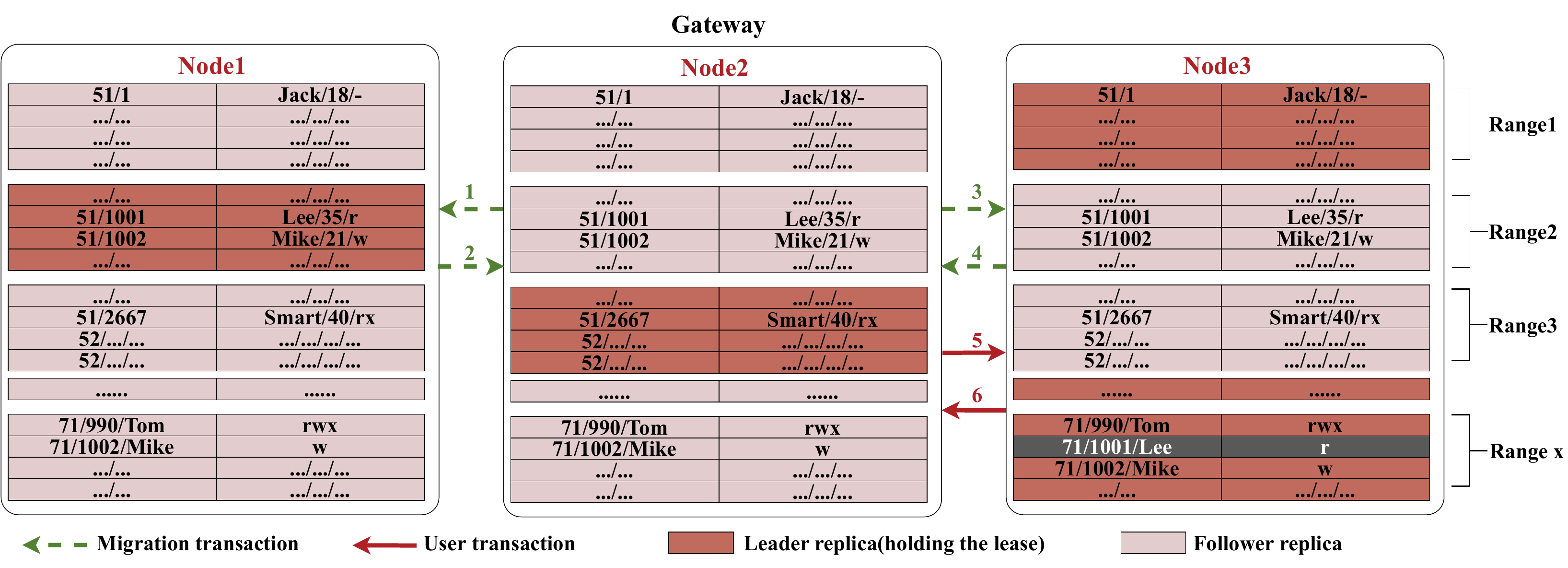}
\caption{Migration and User Transaction Execution}\label{fig4.1}\vspace{-3ex}
\end{figure}

With the basic approach of SLSM, shared-nothing databases can easily perform online schema migrations, avoiding the need to wait for massive data migrations. SLSM is essentially a combination of delayed transaction processing~\cite{faleiro2014lazy} and transaction decomposition~\cite{faleiro2017high,shasha1995transaction,zhang2013transaction}, where large migration transactions are decomposed into separate smaller migration transactions with delayed processing. However, this undoubtedly increases the latency of user transactions on the new schema, as the system is required to generate and execute migration transactions before responding to user transactions. Consider a schema migration of \textbf{table split} from the old \textit{user} table to the new \textit{user\_rights} table shown in Figure~\ref{fig4.0}. The old \textit{user} table contains three slices, each with three replicas, distributed on a three-node cluster, as shown in Figure~\ref{fig4.1}. The old \textit{user} table ID is 51 and the new \textit{user\_rights} table ID is 71. When  processing the user transaction \textsf{SELECT id, rights FROM user\_rights WHERE id $=$ 1001}, gateway node3 needs to first execute the migration transaction \textsf{INSERT INTO user\_rights (id, name, rights) (SELECT id, name, rights from user WHERE id $=$ 1001)}. The relevant data in the old \textit{user} table is on the slice Range2, and the leaseholder of the slice Range2 is node1 at this time. The system routes to node1, obtains the related data, and then migrates to the leaseholder node2 where the corresponding data of the new \textit{user\_rights} table is located. Finally, node3 executes the user transaction. The dashed and solid arrows in Figure~\ref{fig4.1} represent the routing paths of KV operation requests/data for the migration transaction and user transaction, respectively.

Obviously, the additional overhead of migrating transactions causes the latency of user transactions to at least double, and we discuss how to improve the efficiency of migrating transactions and user transactions.
\vspace{-3ex}\subsubsection{Optimization of Migration Transactions.}
We observe that the routing paths for KV operation requests/data of migration transactions are excessive, because the leaseholder of the slice where the source data reside is usually inconsistent with the leaseholder of the slice which is inserted, resulting in additional network round-trip communication overhead. The reason for this phenomenon is two-fold. Firstly, the old and new schemas have different table IDs and primary keys, and the slices in which they reside are different. Secondly, the leaseholder for each slice is dynamically adjusted according to network latency, node load, replica health, and other factors. 

In fact, most schema migration requests are closely related to practical business upgrades. For example, the original table may need to add some new columns to model expanded objects and attributes. As another example, two tables are often joined together and rarely updated independently, and the developer wants to merge them into one to improve query performance. In these cases, the old and new schemas share most of the structure and data. We can leverage this linkage to modify the metadata information of the new schema during the initialization phase so that the leaseholder where the source data slice resides is the same as that of the destination slice in the migration transaction.

\begin{figure}[t]
\centering
\includegraphics[scale=0.22]{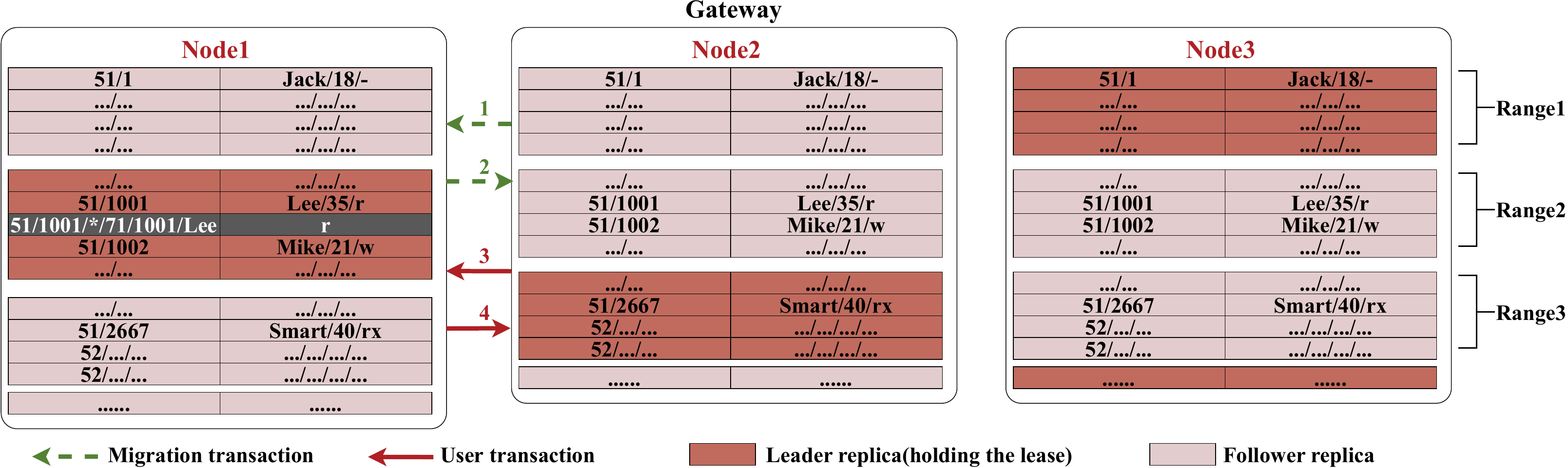}
\caption{Optimization of Migration Transactions}\label{fig4.2}\vspace{-3ex}
\end{figure}

When the new schema contains the primary key of the old schema, SLSM constructs the metadata of the new table in a different way, helping the database system encode and decode the tuples. As shown in Section~\ref{sec:architec}, at the storage layer, the key of key-value pair for the new table is encoded by the new table ID and the primary key columns of the new table, and SLSM adds a ``prefix" to the Key, which consists of the old table ID and the primary key columns of the old table. In other words, the data of the new table are stored with the key of the old table as the prefix. SLSM ensures that the system performs key prefix additions and deletions for subsequent read and write operations to the new table. Consider the schema migration we mentioned in Figure~\ref{fig4.0}, where \textit{user\_rights} table contains the primary key columns of \textit{user} table. As shown in Figure~\ref{fig4.2}, before processing the user transaction \textsf{SELECT id, rights FROM user\_rights WHERE id $=$ 1001}; gateway node3 executes the migration transaction first. The relevant data in the old \textit{user} table is on the slice Range2, and at this time the leaseholder of the slice Range2 is node1. The system routes to node1 and obtains the relevant data, which can then be migrated directly to node1 of the leaseholder in Range2 of \textit{user\_rights} table, reducing the communication overhead compared to Figure~\ref{fig4.1}. It can be seen that the data of the old and new tables are interleaved and that the new schema table has the same data distribution as the old one after all the data have been migrated.

\vspace{-3ex}\subsubsection{Optimization of User Transactions.}
\begin{figure}[t]
\centering
\includegraphics[scale=0.4]{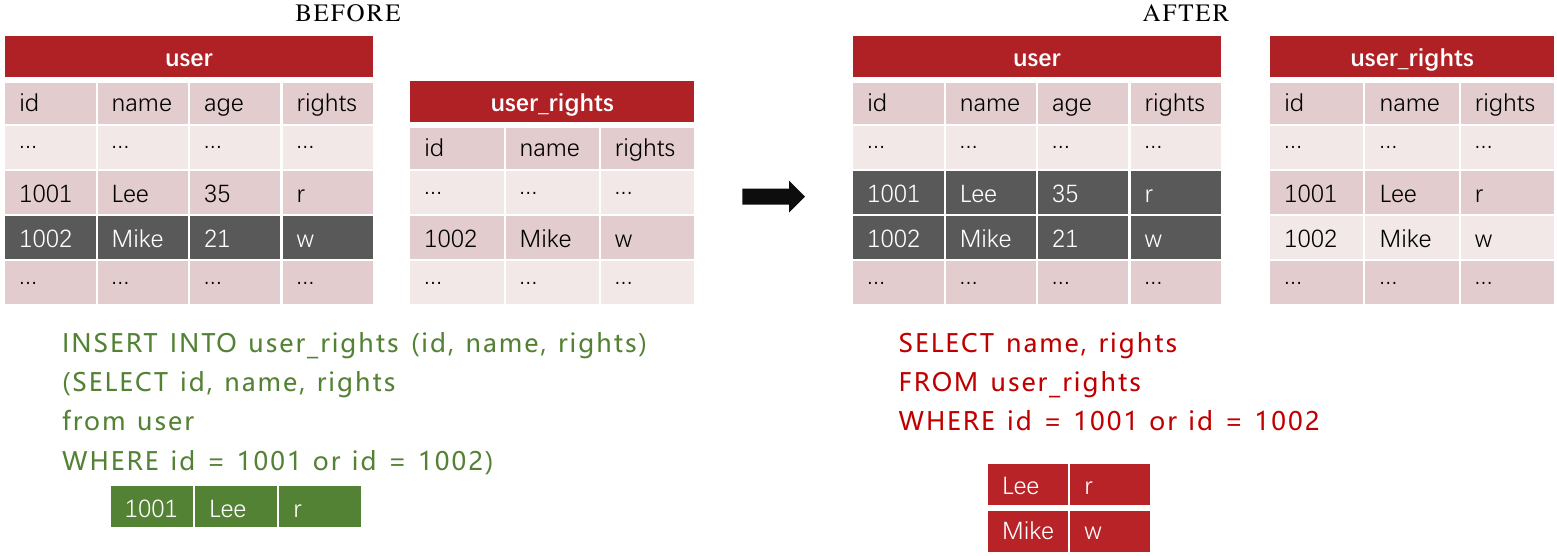}
\caption{Migrating process of migration transaction}\label{appended1}\vspace{-3ex}
\end{figure}

Although we have achieved some performance improvement through migration transaction optimization, it is very limited because user transactions still need to wait for the migration transaction to complete before executing. We observe that the relevant data migrated by the migration transaction are exactly the data needed by the user transaction. The migration transaction is generated by the filter predicate in the user transaction request to ensure that the new schema table has the set of tuples needed to fully process the user transaction. However, due to the concurrency of user transactions, the migration transaction executed before the current user transaction may actually only migrate part of the data, or even not migrate any data, since the migration transaction of other user transactions may have already migrated some or all of the data. As shown in Fig.~\ref{appended1}, the migration transaction (marked in green) is executed before the user transaction (marked in red), and since half of the data has already been migrated (marked in black), the migration transaction only needs to migrate the other half. Therefore, although the data involved in migration transactions are related to user transactions, they may not fully meet the requirements for processing user transactions. However, we can still take advantage of this feature to optimize user transactions, so that the data involved in migration transactions directly serve user transactions, reducing the waiting time for user transactions.

\begin{figure}[t]
\centering
\includegraphics[scale=0.3]{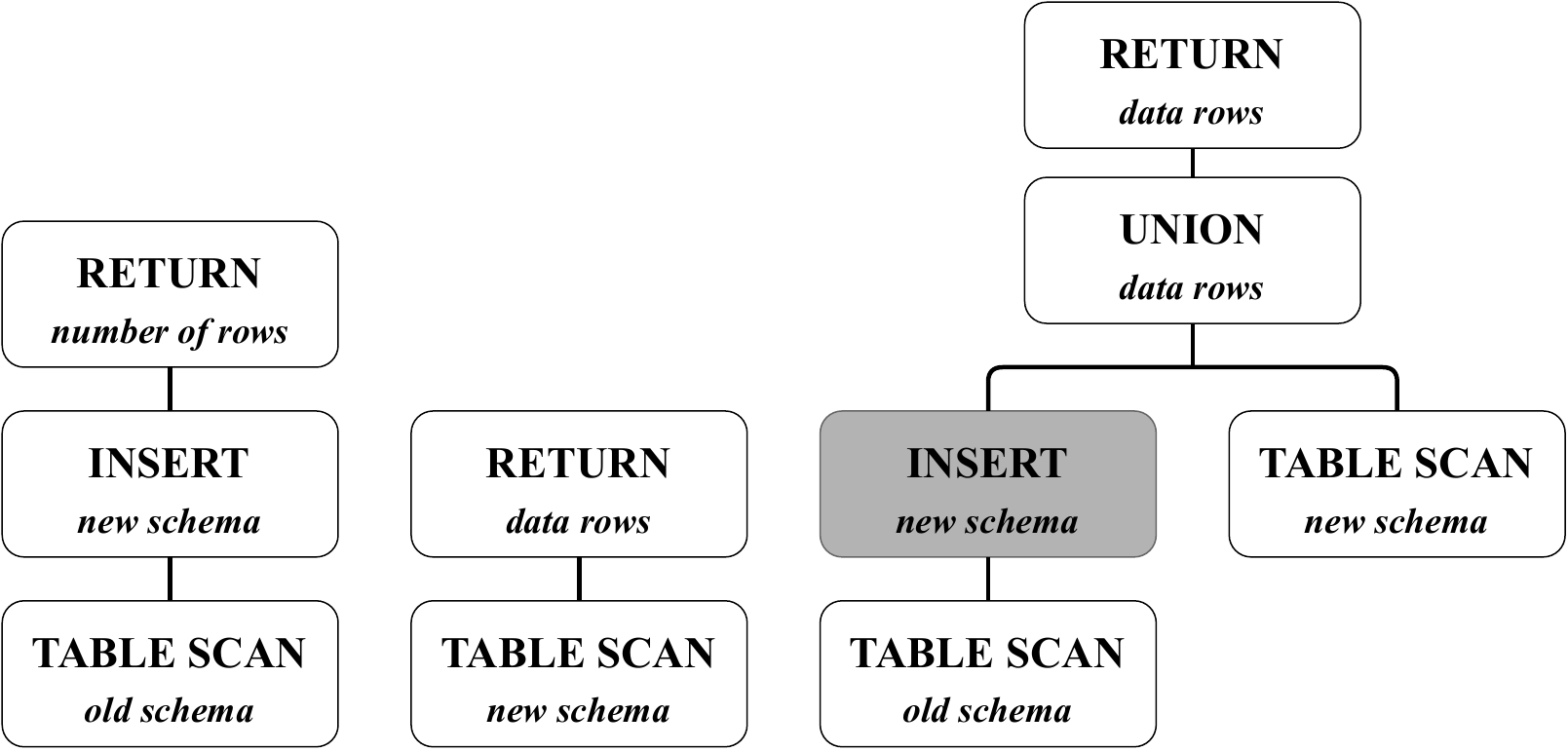}
\caption{Migration, User, and Fusion Transaction for A \textsf{SELECT} Request}\label{fig4.3}\vspace{-2ex}
\end{figure}

\begin{figure}[t]
\centering
\includegraphics[scale=0.21]{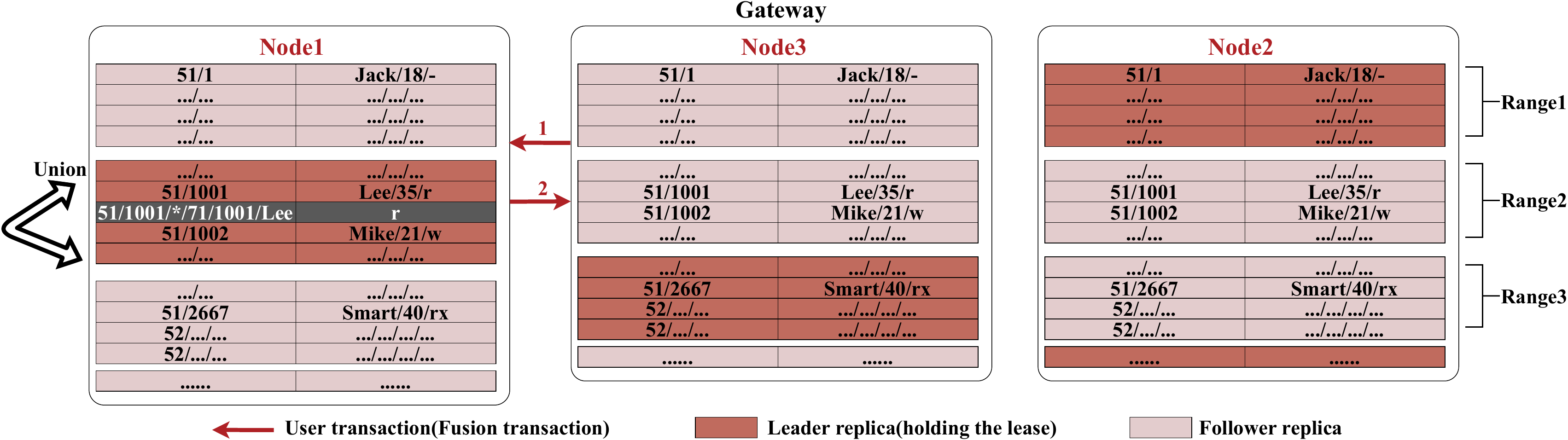}
\caption{Optimization of User Transactions}\label{fig4.4}\vspace{-3ex}
\end{figure}

Instead of strictly generating and executing a migration transaction before a user transaction, SLSM actually executes a user transaction that incorporates the migration transaction (referred to as a fusion transaction). This is achieved by modifying the \textsf{INSERT} operator so that it can output post-insert data in the execution plan operator stream. Figure~\ref{fig4.3} shows the abbreviated execution plans for a migration transaction, a user transaction, and a fusion transaction for a \textsf{SELECT} request, respectively. For \textsf{SELECT} requests, the fusion transaction filters and reads the data from both the old and new schemas. Since the modified \textsf{INSERT} operator returns the inserted data, these data can be merged with the filtered data from the new schema and returned as the result. \textsf{UPDATE} and \textsf{DELETE} requests follow a similar idea, both filter and read the relevant data on the old and new schemas at the same time, and then merge and process them subsequently. The \textsf{INSERT} request is processed directly on the new schema, without the requirement of a fusion transaction. Figure~\ref{fig4.4} shows the advantages of our fusion transaction. When processing user transaction \textsf{SELECT id, rights FROM user\_rights WHERE id $=$ 1001}, the system scans the relevant data on tables \textit{user} and \textit{user\_rights} at the same time in the leaseholder node1 of range2, migrates the data read from \textit{user} table to \textit{user\_rights} table directly, and then merges them with the data read from \textit{user\_rights} table (no relevant data in this case) to return the result.

\vspace{-3ex}\subsubsection{Complexity Analyze.} Finally we study the time complexity of SLSM. We denote the overhead of migration transaction, user transaction and the average communication overhead between nodes as $cost_{mig},cost_{usr}$ and $cost_{com}$, respectively. We categorize the study according to whether the gateway, the old table slice and the new table slice are on the same node. The complexity of the basic SLSM, with migration transaction optimization only, and the final SLSM are shown in Table~\ref{tab1}.

\begin{table}
\centering
\caption{Comparison of complexity of different SLSMs}\label{tab1}
\scalebox{0.7}{ 
\begin{tabular}{|c|c|c|c|}
\hline
Categories &  Basic & Basic-with-mig-trans-optim & Basic-with-all\\
\hline
$\left\{gateway,old,new\right\}$ & $cost_{mig} + cost_{usr}$ &$cost_{mig} + cost_{usr}$&$cost_{mig} + cost_{usr}$\\
$\left\{gateway,old\right\}$ & $cost_{mig} + cost_{usr} + 2*cost_{com}$& $-$ & $-$ \\
$\left\{gateway,new\right\}$ & $cost_{mig} + cost_{usr} + 2*cost_{com}$ & $-$& $-$ \\
$\left\{old,new\right\}$ & $cost_{mig} + cost_{usr} + 2*cost_{com}$ & $cost_{mig} + cost_{usr} + 2*cost_{com}$& $\left\{cost_{mig},cost_{usr}\right\}_{max} + 2*cost_{com}$  \\
$\varnothing$ & $cost_{mig} + cost_{usr} + 6*cost_{com}$ & $-$& $-$ \\
\hline
\end{tabular}
}
\end{table}

\vspace{-2ex}\section{Experiments}\label{sec:exp}
\subsection{Experimental Setup}
We conducted our experiments on three AliCloud server instances, each with 4 vCPUs at 2.70 GHz on Intel(R) Xeon(R) Platinum and 8 GB of RAM.
\vspace{-3ex}\subsubsection{Implementation.}
We implemented a prototype of SLSM on top of CockroachDB 21.1.21. CockroachDB is an open source distributed NewSQL database engine, following the shared-nothing architecture. We built a three-node CockroachDB cluster. By default, the average round-trip time of the nodes in the cluster is around 11.7ms (set via the``\textsf{tc}" command under Ubuntu).
\vspace{-3ex}\subsubsection{Baselines.}
We compare the performance using the following approaches:
\begin{itemize}
    \item[i)] \textbf{Upperbound:} Vanilla CockroachDB cluster without any schema migration. We use it to show the upper bound.
    \item[ii)] \textbf{OSC:} CockroachDB’s own online schema change schemes~\cite{taft2020cockroachdb} built upon work originated by the F1 team at Google. It backfills (or deletes) the underlying table data without locking them and thus without any downtime.
    \item[iii)] \textbf{Bullfrog:} The state-of-the-art lazy schema migration method~\cite{bhattacherjee2021bullfrog} in a standalone database, implemented in Postgresql 11.0. We port it to CockroachDB for experimental comparison.
    \item[iv)] \textbf{SLSM:} Our lazy schema migration solution specifically designed for shared-nothing databases.
\end{itemize}
\vspace{-3ex}\subsubsection{Benchmarks.}
To evaluate SLSM under real OLTP workloads, we, therefore, follow previous work~\cite{bhattacherjee2021bullfrog} to use a variation of TPC-C that includes schema migrations. By default, we use a scale factor of 50. We start the benchmark by running the original TPC-C mix, and after some seconds, we start a schema migration to perform one of the following operations:
\begin{itemize}
    \item[1)] \textbf{SplitTable:} Split the \textit{customer} table into two, one containing private customer information such as credit, payment and balance, and the other containing public customer information like state, city, street, etc. The two new tables after the split have the same primary key as the original \textit{customer} table.
    \item[2)] \textbf{JoinTable:} Join the \textit{stock} and \textit{order\_line} tables. This optimizes the StockLevel transaction, which reads stock after scanning the \textit{order\_line} table to get out-of-stock items.
    \item[3)] \textbf{Preaggregate:} Sum up the values in \textit{order\_line} table where ol\_w\_id = o\_w\_id, ol\_d\_id = o\_d\_id and ol\_o\_id = o\_id. The results are maintained as a separate table.
\end{itemize}

\vspace{-3ex}\subsubsection{Experimental Platform and Metrics.}
We use BenchBase~\cite{difallah2013oltp} to setup and run our experiments. BenchBase supports tight control of transaction mixtures, request rates, and access distributions over time. We measure throughput as transactions per second (TPS) and the end-to-end latency as the time from when the client issues a transaction request until the response is received. The measurements for all of our experiments are averaged over 5 runs, but we found that the variance across runs in each of our experiments was negligible.

\vspace{-3ex}\subsection{Performance under Schema Migration}
\begin{figure}[t]
\centering
\includegraphics[scale=0.45]{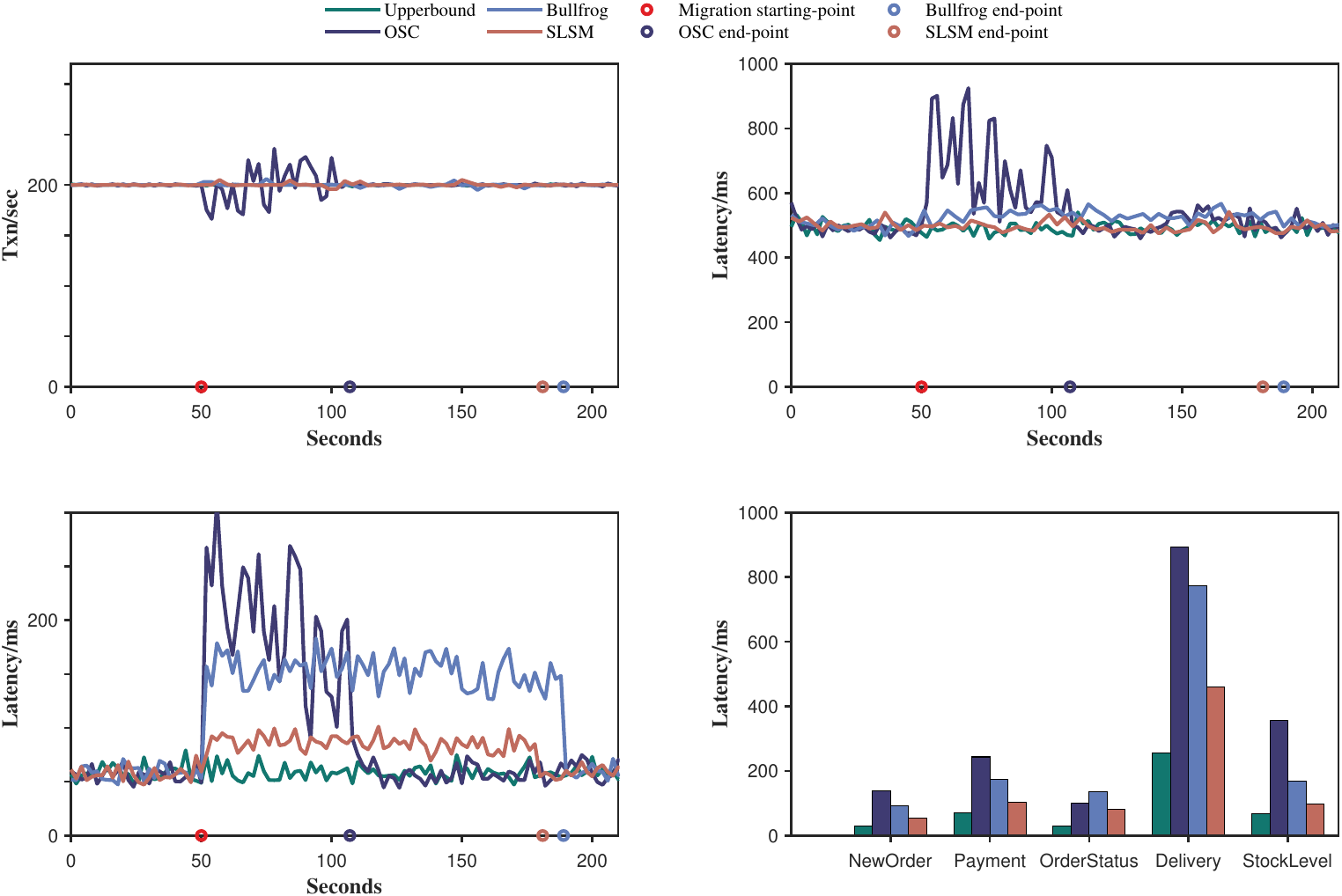}
\caption{Throughput and latency during migration} \label{fig5.1}
\end{figure}

Our first set of experiments focus on how the throughput and latency of transaction processing vary during the different phases of schema migration. Once the schema migration request starts, for Buffrog and SLSM, transactions containing requests on the old schema are immediately replaced by transactions containing requests on the new schema (we have modified them to be compatible); for OSC, old transactions are still used until the migration is complete. Fig.~\ref{fig5.1} illustrates our experimental results. The migration begins for all implementations at the red circle and ends for each system at the later corresponding circles marked in the figure.

Consider the upper two subplots, OSC takes approximately 60 seconds to complete, and transactions experienced significant jitter in both throughput and latency during the migration. Although access to the old schema is not blocked during the migration, data backfilling and double-writing of the old and new schemas cause some negative impact on system performance. No noticeable fluctuations in throughput or latency are observed for Bullfrog or SLSM, and the total time to complete the migration is longer because the migration will take place on demand while the user transactions are being processed. Compared to the background migration algorithm in Bullfrog, SLSM limits simulated user transactions to tuples that have not yet been migrated, rather than all tuples for the entire table, and therefore has a shorter migration time window. Although SLSM has better latency performance, it is not significant compared to Bullfrog. Since each transaction in the TPCC benchmark involves requests on many different tables, there may be only two or three requests on the new schema. Therefore, we simplify each type of transaction in TPCC by only evaluating the latency of those requests on the new schema. As shown in the lower left subfigure, the latency of SLSM improves by about 40\% overall compared to Bullfrog, and we illustrate the latency improvement for each specific type of transaction in the lower right subfigure. 

\vspace{-3ex}\subsection{Ablation Experiments}
\begin{figure}[t]
\centering
\includegraphics[scale=0.47]{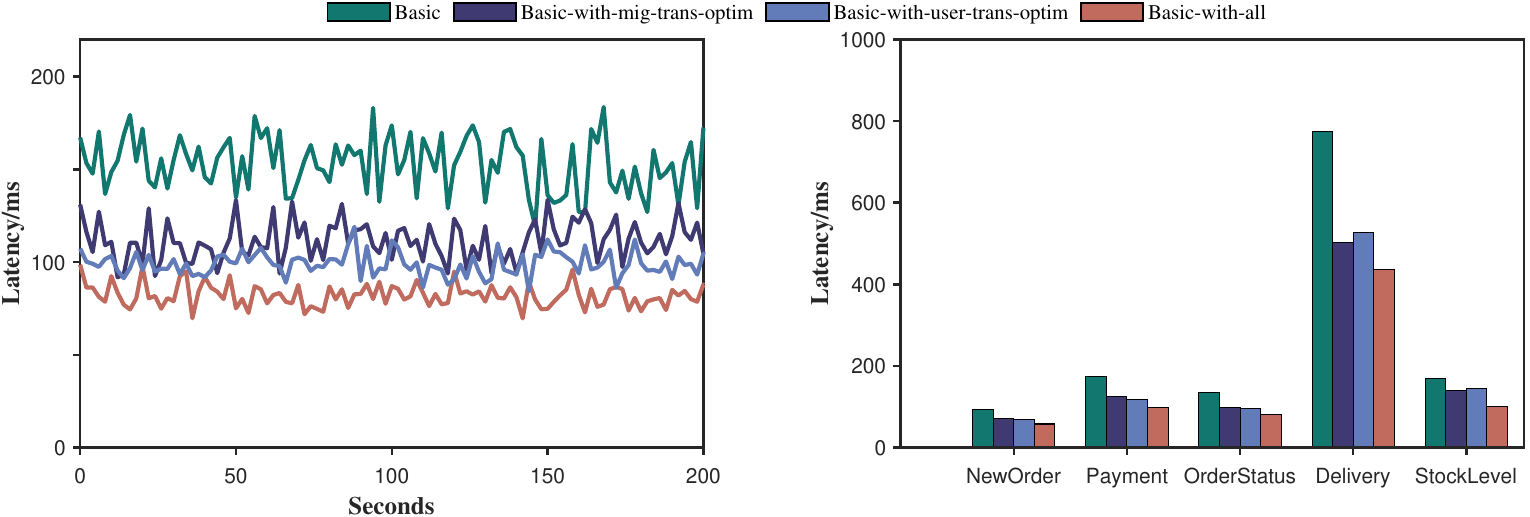}
\caption{Latency during migration under different SLSMs} \label{fig5.2}
\end{figure}

We then evaluate the optimization of migration transactions and user transactions in SLSM (see Section~\ref{sec:furOpt}). For visual comparison, we start schema migration directly and stop the background migration process, using TPCC transaction (evaluating requests on the new schema only). As shown in Fig.~\ref{fig5.2}, we label the basic SLSM, the basic SLSM with migration transactions optimization only, the basic SLSM with user transactions optimization only, and the final SLSM as ``Basic", ``Basic-with-mig-trans-optim", ``Basic-with-user-trans-optim", and ``Basic-with-all". It turns out that each of our optimization methods performs as expected. Either optimization alone reduces the latency of user transactions in the basic SLSM, while combining the two optimization methods allows SLSM to achieve the best performance.
\vspace{-3ex}\subsection{Impact of Network Round-Trip Time}
\begin{figure}[t]
\centering
\includegraphics[scale=0.42]{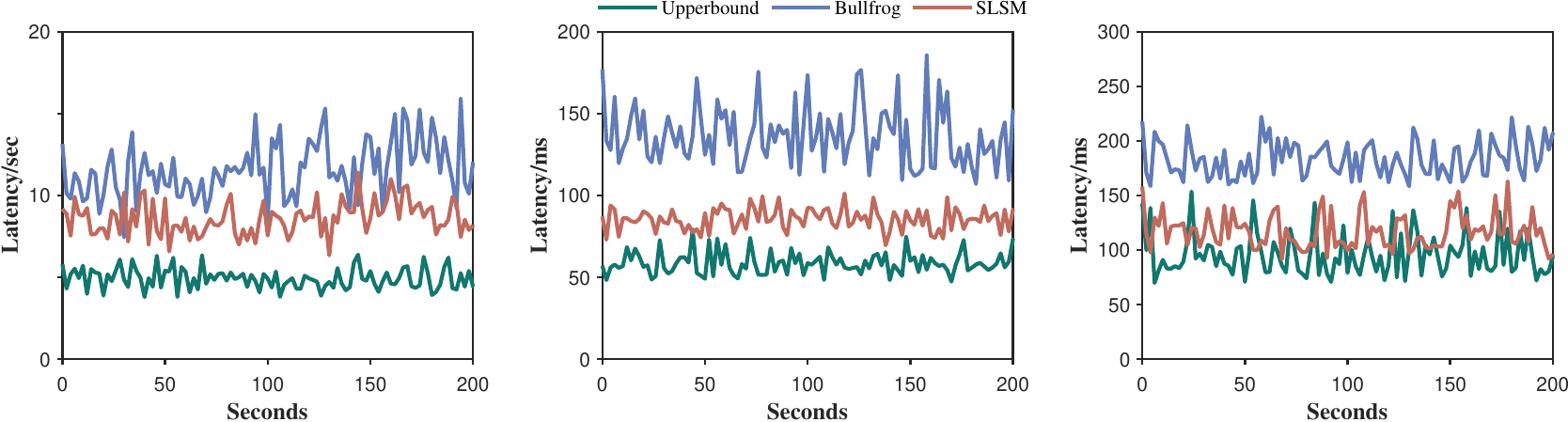}
\caption{Latency during migration with different network round-trip time} \label{fig5.3}
\end{figure}

Now we examine how network round-trip time (RTT) affects schema migration performance. Network RTT is an important metric for distributed databases, as data are often distributed between multiple nodes or locations, which means that data operations (\eg read, write, and transactions) involve network communication. Our experimental setup is similar to the above, with the addition of measurements for the Upperbound and Bullfrog. As shown in Fig.~\ref{fig5.3}, the average RTT in the experimental results from left to right are 1.15ms, 11.78ms and 22.33ms, respectively. We observe that as the RTT grows, the latency of SLSM converges more towards the Upperbound. As network communication overhead gradually becomes the bottleneck of transaction execution performance, since migration transaction optimization and user transaction optimization in SLSM can reduce the network communication overhead to a certain extent, it has a latency curve closer to Upperbound compared to Bullfrog. At low RTT, the reduction in network communication latency in migration transaction optimization by SLSM is not significant compared to its extra overhead in the data's key. However, the optimization of user transactions still offers an advantage to SLSM.

\vspace{-3ex}\subsection{Stand-alone and Clustered}
\begin{figure}[t]
\centering
\includegraphics[scale=0.45]{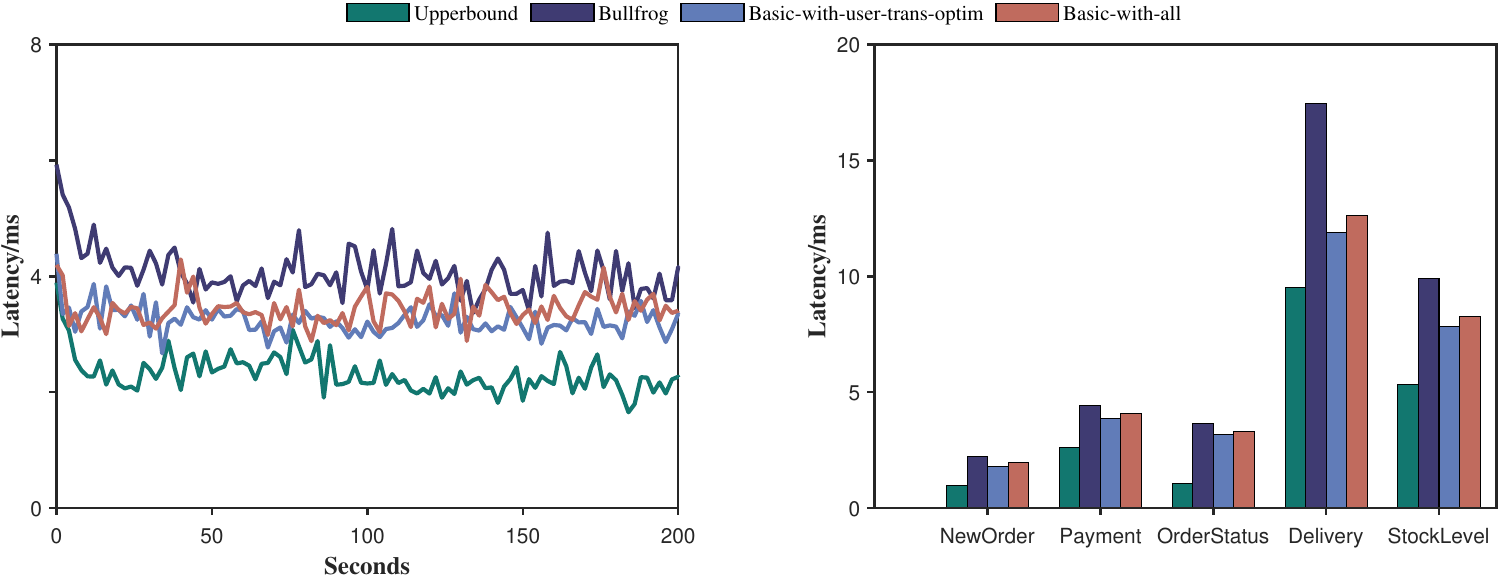}
\caption{Latency during migration in a single-node cluster} \label{fig5.4}
\end{figure}

Although SLSM is designed for distributed shared-nothing databases, it is also applicable on stand-alone databases. In the final experiment, we analyze SLSM in a single node cluster; the results are shown in Fig.~\ref{fig5.4}. In addition to Bullfrog and SLSM, we also compare the basic SLSM with user transaction optimization only since migration transaction optimization designed specifically for network communication overheads no longer works. It turns out that SLSM still has a performance advantage over the current state-of-the-art standalone database lazy schema migration scheme and performs better with the migration transaction optimization removed. Migration transaction optimization struggles more in terms of applicability than user transaction optimization. Specifically, the former optimizes network communication overhead and, more importantly, effectively reduces the time that user transactions wait for execution. For standalone databases that do not require a network for node communication, the work we did with user transaction optimization in SLSM still paid off.

\vspace{-3ex}\section{Conclusion}\label{sec:concl}
Schema migrations on shared-nothing databases typically last a long time, since they are accompanied by massive data movement, resulting in a service vacuum before the new schema is available. In this paper, we propose SLSM, a lazy migration strategy, to perform online schema changes so that the new schema can be immediately ready for access, even when the physical data has not yet been migrated to the new schema. SLSM improves the performance of user transactions during migration by decreasing the network communication overhead and the waiting time for user transactions. Experimental results indicate that SLSM can accomplish schema migration on shared-nothing databases with high quality and with low impact on user transaction latency. The solution not only works on shared-nothing databases, but is also applicable for stand-alone database systems.
\begin{credits}
\subsubsection{\ackname} This work is supported by the Natural Science Basic Research Program of Shaanxi under Grant No.2023-JC-QN-0648, the National Natural Science Foundation of China under Grant No.62302370.
\end{credits}

\bibliographystyle{splncs04}
\bibliography{sample-base}
\end{document}